\documentclass[aps,prl,twocolumn,groupedaddress]{revtex4}

\usepackage{bm} 
\usepackage{graphicx}

\def\Vec#1{\mbox{\bm{$#1$}}}

\begin{document}


\title{Adiabatic Loading of Cold Bosons in Three-Dimensional Optical Lattices and Superfluid-Normal Phase Transition}


\author{S.~Yoshimura$^{1,2}$, S.~Konabe$^3$, and T.~Nikuni$^3$}
\affiliation{$^1$Department of Physics, Graduate School of Science, The
University of Tokyo, 7-3-1 Hongo, Bunkyo-Ku, Tokyo 113-8656, Japan\\
$^2$CREST, JST, 4-1-8 Honcho Kawaguchi, Saitama 332-0012, Japan\\$^3$Department of Physics, Faculty of Science, Tokyo
University of Science, 1-3 Kagurazaka, Shinjuku-ku, Tokyo 162-8601, Japan}


\date{\today}

\begin{abstract}
We investigate the effects of the adiabatic loading of optical lattices
 to the temperature by applying the mean-field approximation to the
 three-dimensional Bose-Hubbard model at finite temperatures. We
 compute the lattice-height dependence of the isentropic curves for the given initial temperatures in case of
the homogeneous system i.e., neglecting the trapping potential. Taking the unit
 of temperatures as the recoil energy, the adiabatic
 cooling/heating through superfluid (SF) - normal (N) phase transition is clearly understood. It is found that the cooling occurs in
 SF phase while the heating occurs in
 N phase and
 the efficiency of adibatic cooling/heating is higher at higher temperatures. We
also explain how its behavior can be understood from the lattice-hight
 dependence of dispersion
 relation in each phase. Furthermore, the connection of the adiabatic
 heating/cooling between the cases with/without the trapping potential is discussed.
\end{abstract}
\maketitle
Recently ultracold atoms in optical lattices have been studied
intensively both theoretically and experimentally (for the reviews, see
\cite{jaksch2005cah,morsch2006dbe,lewenstein2007,2007arXiv07043011B}).
Not only atomic, molecule, and optical (AMO) physics party, but also
quantum information and condensed matter physics party come into
this field and it has the possibilities of producing new kind of physics. From
the viewpoint of quantum information, ultracold atoms in optical
lattices can be used as one-way quantum computing
\cite{Clark05,Kay06,Christandl05} and dynamical controlling of entanglement
\cite{Jaksch99,mandel2003ccm} under well-controlled conditions. Considering strongly correlated physics,
this system offers the possibility of realizing
various quantum lattice models like the Bose-Hubbard model, the Fermi-Hubbard
model, and the Bose-Fermi Hubbard model, which have various rich quantum
phases \cite{lewenstein2007}.

In order to investigate the above subjects, it is crucial to understand the
lattice-height dependence of the temperatures. In the experiments, the temperature of Bose gases is measured before inserting
optical lattices. However, the experimental method to investigate the
temperature of Bose gases in optical lattices has not been established. Usually the loading
process can be treated as adiabatic since the loading speed is very
low. The
behavior of the temperature of this system during the adiabatic loading of optical
lattices is therefore of great interests, which has been studied in
several papers \cite{blakie2004alb,rey2006rit,ho2007,Gerbier2007}. 

For the non-interacting Bose gases, the adiabatic cooling
only occurs in the tight-binding regime (on the other hand, the adiabatic
heating occurs when the thermal energy lies in the first excited
band). The mechanism of the adiabatic heating/cooling in this case can be understood in
terms of the
change of density of states \cite{blakie2004alb}. The adiabatic loading
including the interaction
between atoms and the effect of the trapping potential has been studied for deep lattices. In
that case, it is found that the adiabatic heating occurs due to increasing
the Mott gap and the trapping effect induces the adiabatic compression and
expansion which cause the adiabatic heating and cooling
\cite{rey2006rit,ho2007,Gerbier2007}. On the other hand, the mechanism
of the adiabatic cooling/heating through SF-N phase transition is not fully understood. 

In this paper, we examine the three-dimensional Bose-Hubbard model at finite temperatures
within the mean-field approximation in order to investigate the effect of
the adiabatic loading to the temperature of the system through SF phase to N
phase. We show that the adiabatic cooling occurs in SF phase while
 the adiabatic heating occurs in N phase. The number fluctuation conserves
 during the adiabatic loading after crossing SF-N phase transition point,
 which yields that it is necessary to have ultracold temperatures at
the phase transition point in order to obtain the system with very low
 number fluctuation at deep optical lattices. We argue that the mechanism
 of the adiabatic cooling/heating is due to the dispersion relation in
 each phase. Finally, we will mention that in the case with the trapping
 potential, the mechanism of the adiabatic heating/cooling is
 essentially same as in the case without the trapping potential.

The Hamiltonian for Bose atoms in optical lattices can be written as
\begin{eqnarray}
\hat{H} &=& \int d^3 x \hat{\psi} ^{\dagger} (\Vec{x})
 \left[-\frac{\hbar ^2}{2 m} \nabla ^2 + V_{o} (\Vec{x}) \right] \hat{\psi} (\Vec{x}) \nonumber \\
&+& \frac{1}{2} \frac{4 \pi a_s
 \hbar ^2}{m} \int d^3 x  \hat{\psi} ^{\dagger} (\Vec{x})
 \hat{\psi} ^{\dagger} (\Vec{x}) \hat{\psi} (\Vec{x}) \hat{\psi}
 (\Vec{x}) \label{microscopicHamiltonian}
\end{eqnarray}
where $\hat{\psi} (\Vec{x})$ is a field operator for Bose atoms and 
$V_{o} (\Vec{x})$ is the optical lattice potential. We consider
three-dimensional optical lattices where $V_{o} (\Vec{x})$ has the form
\begin{equation}
V_{o} (\Vec{x}) = V (\sin ^2 k x + \sin ^2 k y +\sin ^2 k z).
\end{equation}
Here $k = 2 \pi /\lambda$ where $\lambda$ is the wave length of
standing wave laser forming optical lattices. The lattice constant is determined by $a = \lambda
/2$. The lattice height of optical lattices $V$ is measured by the recoil
energy $E_R = \hbar ^2 k^2 /2 m$ where $m$ is mass of the atom. We use
the dimensionless lattice height $s =
V /E_R$. The binary interaction between atoms is approximated by s-wave
scattering, which is charactrized by the scattering length $a_s$.

The Bose-Hubbard Hamiltonian can be
derived by applying the tight-binding approximation to
Eq.~({\ref{microscopicHamiltonian}}) \cite{Jaksch98}. We expand the field operator by the
Wannier function $w_0 (\Vec{x}-\Vec{x}_i)$ as $\hat{\psi} ^{\dagger} (\Vec{x}) = \sum _i \hat{b_i}^{\dagger} w
_0 (\Vec{x}-\Vec{x}_i)$, where $\hat{b_i}$ is the destruction operator
for a boson at a lattice site $\Vec{x}_i$. We can rewrite Eq.~({\ref{microscopicHamiltonian}})
as
\begin{equation}
\hat{H} = -t \sum _{\langle ij \rangle} (\hat{b}_i ^{\dagger} \hat{b_j}+h.c.) +
 \frac{U}{2} \sum _i \hat{n}_i (\hat{n}_i-1) -\sum _i \mu \hat{n}_i .\label{H}
\end{equation}
As usual, we consider the nearest-neighbor hopping and the on-site
interaction. Note that we added the chemical potential $\mu$ in order to treat the
grand-canonical ensemble.
The relation between $U,t$ and $s$ under approximating the Wannier function
as the Gaussian functioin can be found as
$
 \frac{U}{E_R} = 5.97 \frac{a_s}{\lambda} s^{0.88}$ and 
$ \frac{t}{E_R} = 1.43 s^{0.98} e ^{-2.07 \sqrt s}
$
\cite{Gerbier05}.
In this paper, we consider $^{87}$Rb and set $m=1.44 \times 10^{-25}$ [kg], $a_s = 545$ [nm], and $\lambda = 852$
[nm] as in the Greiner's experiment \cite{Greiner02}.

The validity of using the Bose-Hubbard model was examined in
Ref.~\cite{Jaksch98}. The tight-binding approximation at finite
temperatures implies that the
energy scale including the thermal energy must be less than the gap energy
between the lowest band and the second band. We calculated the thermal
energy $k_B T$ normalized by the gap energy $\Delta$ and confirmed $k_B
T/\Delta \ll 1$ for the parameters we will use in this paper.

\begin{figure}
  \begin{center}
    \begin{tabular}{cc}
      \resizebox{45mm}{!}{\includegraphics{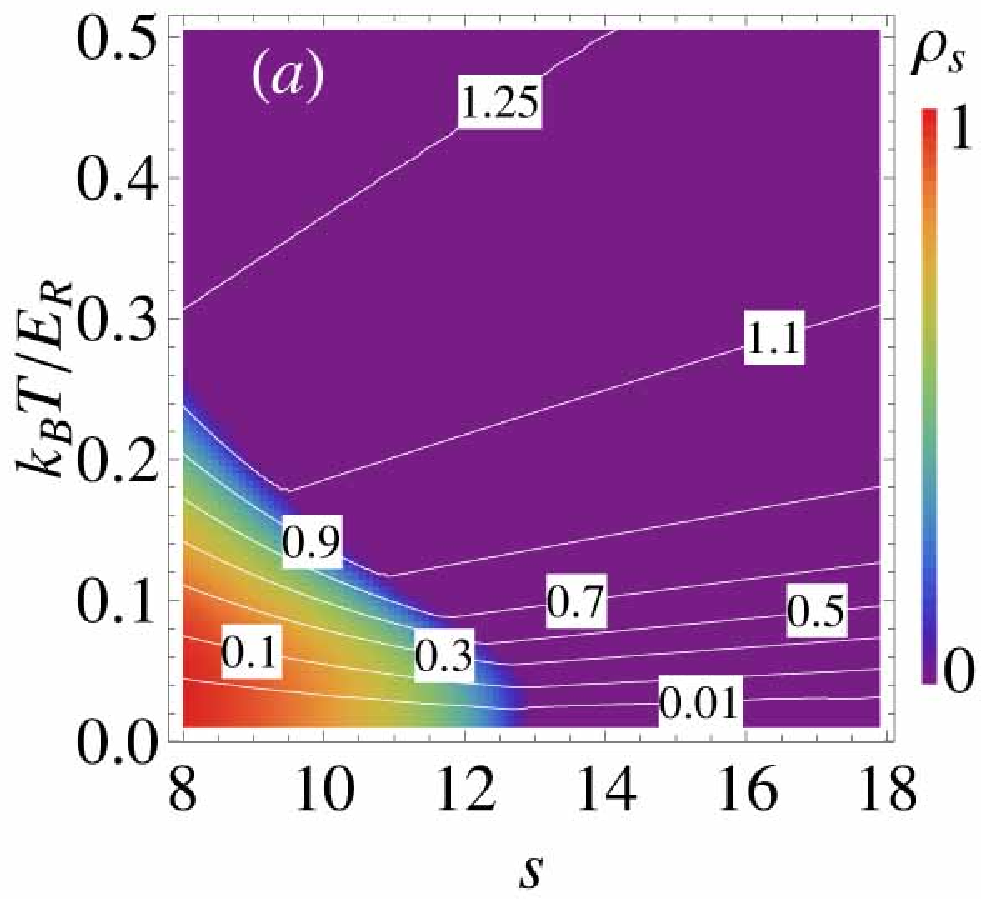}} &
      \resizebox{45mm}{!}{\includegraphics{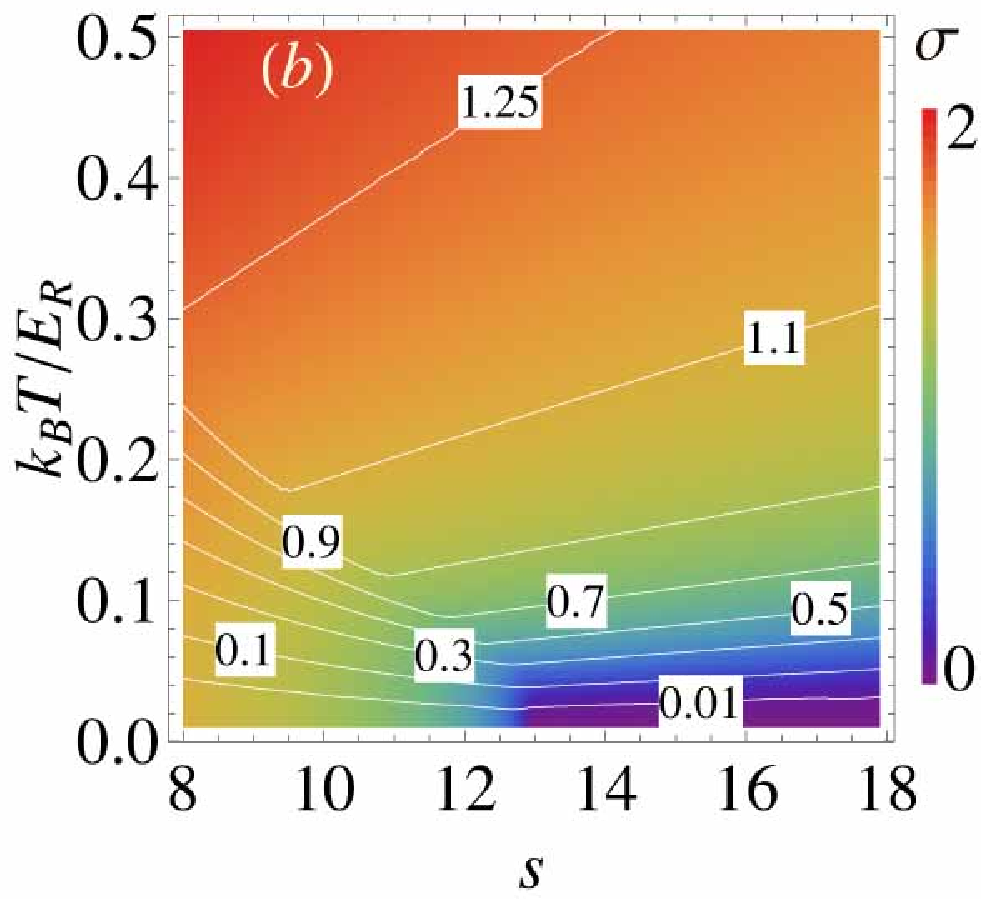}} \\
    \end{tabular}
    \caption{(Color online) (a) The isentropic curves and the condensate density $\rho _s$ for
   the given lattice height $s$ and temperatures $k_B T/E_R$. We choose
   the chemical potential $\mu$ which
   satisfies $\rho = 1$. The number inside box on the isentropic curve indicates the
 value of the entropy $S/k_B$. The system goes along the isentropic curve
   which corresponds to the initial entropy. SF phase stands for the cooling
   region, whereas N phase stands for the heating region. (b)  
The isentropic curves added the density plot of the number fluctuation
   $\sigma$. The isentropic curves is the same one as in (a). In SF phase, the number fluctuation is
   enhanced due to the quantum fluctuation. As loading optical lattices
   adiabatically, the number fluctuation decreases and it takes the minimum
   value at
   the critical lattice height $s_c$. The temperature increases after
   crosing $s_c$ whereas the number fluctuation is same.
}
    \label{Adiabatic}
  \end{center}
\end{figure}

Let us apply the mean-field approximation \cite{Oosten01,MFAFiniteT} as $\hat{b} ^{\dagger} _i
\hat{b}_j \approx \phi (\hat{b} ^{\dagger} _i + \hat{b}_j) - \phi ^2$
where $\phi = \langle \hat{b} ^{\dagger} \rangle = \langle \hat{b}
\rangle$ is taken to be real. Then we can rewrite the Hamiltonian
Eq.({\ref{H}}) as
sum of on-site Hamiltonians as $\hat{H} = \sum _i \hat{H_i}$, and the Hamiltonian of
the $i$-th site is 
\begin{equation}
\hat{H}_i = \frac{U}{2} \hat{n_i} (\hat{n_i}-1) -zt \phi
 (\hat{b}^{\dagger}_i+\hat{b}_i)+zt\phi ^2 -\mu \hat{n}_i.\label{MeanFieldBoseHubbard}
\end{equation}
Here $z$ denotes the coordinate number. In the actual conditions, we
truncate the 
size of the Hilbert space of $\hat{H}_i$ by assuming the maximum number of particles which
can be at one site is $n_t$. We take large $n_t$ so that the truncation
effect on the calculated physical quantities is neglegible. Diagonalizing this
Hamiltonian $\hat{H}_i$ under given $U$, $t$, and $\mu$, we obtain the eigenstates and
the corresponding eigenenergies. Then we can calculate the partition function $Z$ and
the Helmholtz free energy $F$ as functions of $\phi$ for given
temperatures. $\phi$ is determined by the self-consistent equation $\partial
F / \partial \phi =0$.
At first, we assume that the system is homogeneous i.e., neglecting the trapping potential. In order to investigate the behavior of this system
during the adiabatic
loading of optical lattices, we calculate the entropy $S/k_B$, the condensate
density $\rho _s \equiv \phi ^2$, and
the number fluctuation $\sigma \equiv \sqrt {\langle \hat{n} ^2 \rangle - \langle
\hat{n} \rangle ^2}$ under the fixed chemical potential $\mu$ as the
occupation number $\rho = 1$. Fig.~{\ref{Adiabatic}} (a) shows the relation
between the isentropic curves and the condensate density. Loading optical lattices adiabatically, the system goes along the isentropic curve which
corresponds to the initial entropy. One finds that there exists two
areas: the adiabatic cooling region and the adiabatic heating one. Former is the region where the system
is cooled as the lattice height increases while the latter is that of where
the system is heated as the lattice height increases. Notice that the
efficiency of the adiabatic cooling/heating is higher for higher
initial entropy. Seeing the condensate density, one finds that SF phase,
which is charcterized by $\rho _s >0$, exists in shallow lattices and
at low temperatures. The cooling
region lies in SF phase, whereas the heating region lies in N phase where
$\rho _s=0$. We also find that the efficiency of the adiabatic cooling/heating is not
related with the amount of the condensate density.


We plot the number fluctuation and the isentropic curves in
Fig.~{\ref{Adiabatic}} (b). For $k_B T/E_R < 0.04$, there exists a
rectangle region with the very low number fluctuation, which is called as
the thermal insulator in Ref.~\cite{Gerbier07}. In that region, the critical lattice height $s_c$ does
not change much. It is necessary to satisfy $S/k_B < 0.01$ in order to
reach the thermal insulator. Note, however, that the number fluctuation is
constant on the same isentropic curve in N phase. Thus, although the
system is heated up, the number fluctuation does not change with loading
optical lattices adiabatically after crossing SF-N phase transition point.

\begin{figure}
  \begin{center}
    \begin{tabular}{cc}
      \resizebox{42mm}{!}{\includegraphics{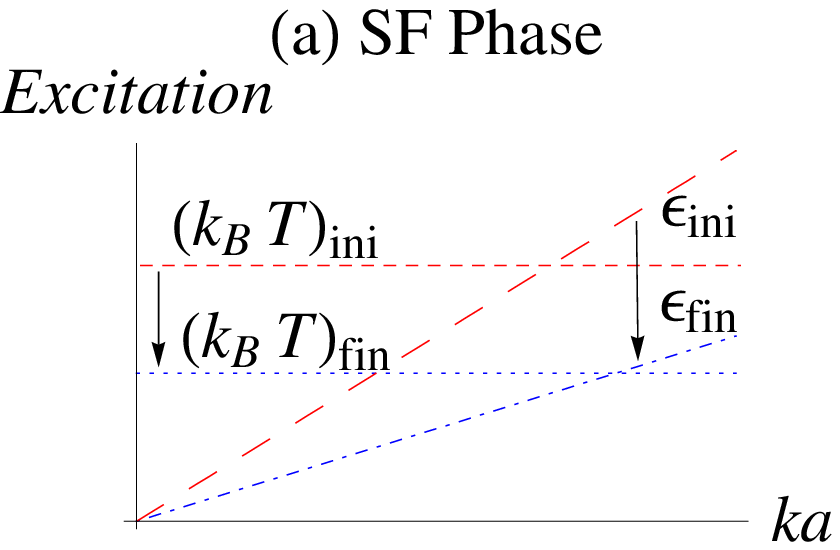}} &
      \resizebox{42mm}{!}{\includegraphics{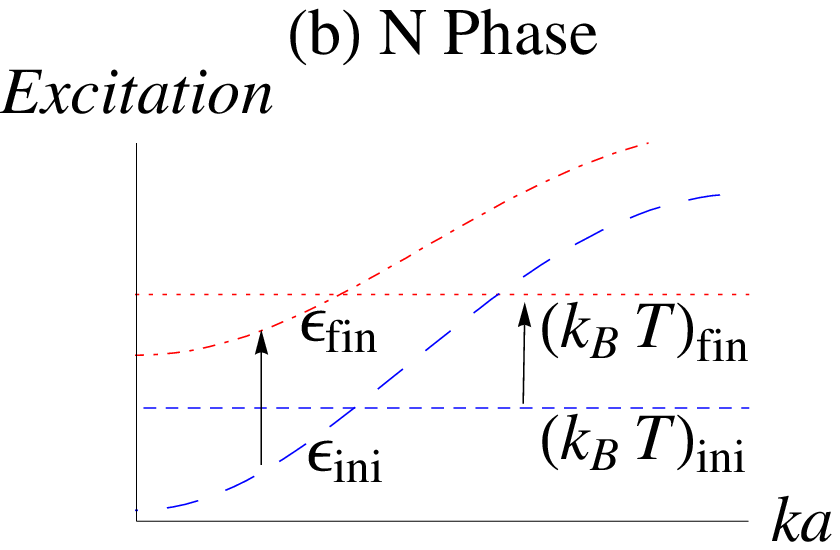}} \\
    \end{tabular}
    \caption{(Color online) Schematic picture of the dispersion relation
   and the thermal
   energy. Arrows indicate the direction of loading optical
   lattices. Its tail (head) corresponds to the initial (finial)
   state. (Dot: the thermal energy in the initial state $(k_B T)_{\rm ini}$, Short Dashed:
   the thermal energy in the final state $(k_B T)_{\rm fin}$, Long Dashed: The excitation spectrum
   in the initial state $\epsilon _{\rm ini}$, Dot Dashed: The excitation
   spectrum in the final state $\epsilon _{\rm fin}$) (a) SF phase. The sound
velocity decreases with increasing the lattice height. Thus the temperature must
   decrease in order to keep the number of state. (b) N phase. The energy
   gap opens with loading optical lattices. Hence the temperature increase
   to maintain the entropy being a constant.  
}
    \label{dispersion}
  \end{center}
\end{figure}

One can understand the reason why the adiabatic cooling occurs in SF phase while the adiabatic
heating does in N phase by investigating the
dispersion relation in each phase (see Fig.~{\ref{dispersion}}). The adiabatic loading
means that the number of state is conserved during this process. As
shown in Fig.~{\ref{dispersion}} (a), the dispersion relation in SF phase
exhibits
the phonon-dispersion, where the sound
velocity decreases with increasing the lattice height from the initial
state. Then the number of
state which can be occupied by the thermal energy increases if the temperature does not
change. Thus the temperature must decrease in the final state in order to keep the number of state
(i.e. the entropy)
being constant. Fig.~{\ref{dispersion}} (b) shows that, in N phase, the
energy gap increases as the lattice height increases from the initial
state. Then the
number of state which can be accesjsible by the thermal energy
decreases if the temperature is same. Therefore the temperature rises to
maintain the number of state being constant in the final state. In this way, the behavior
of the adiabatic
heating/cooling is determined by whether the dispersion relation increases
or decreases from the initial state to the final state.

\begin{figure}[htbp]
  \begin{center}
    \begin{tabular}{cc}
      \resizebox{45mm}{!}{\includegraphics{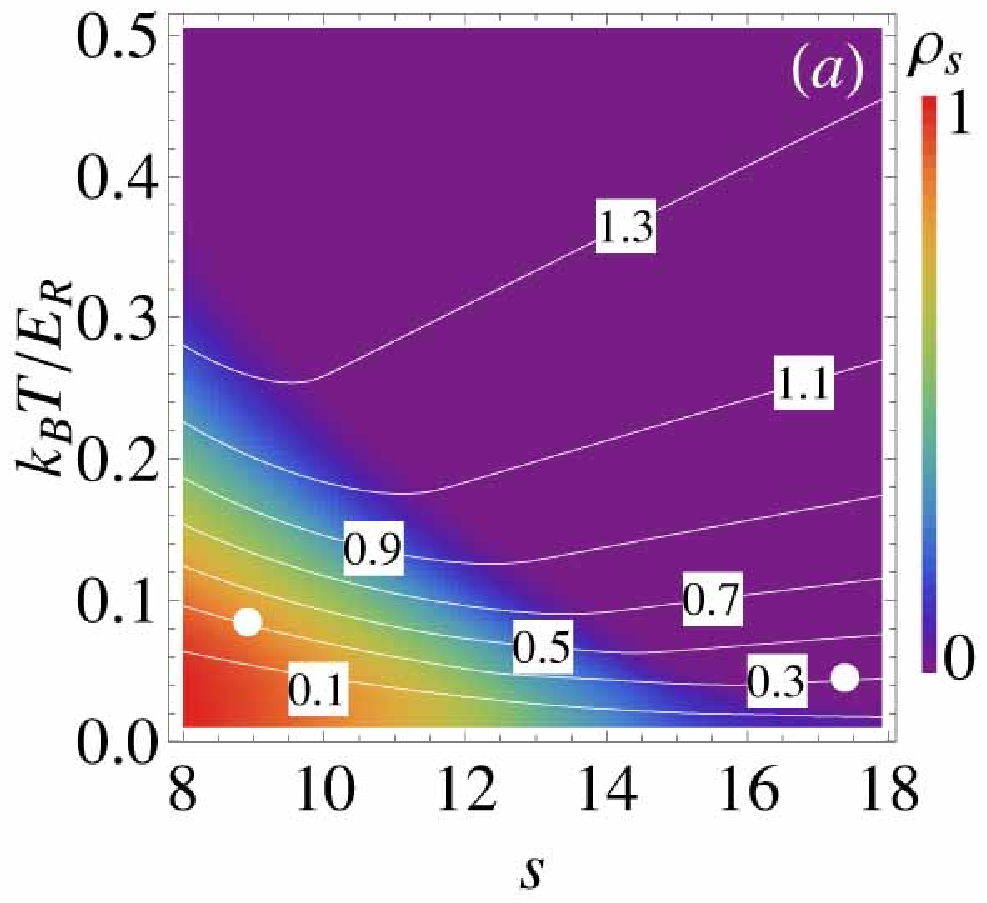}} &
      \resizebox{45mm}{!}{\includegraphics{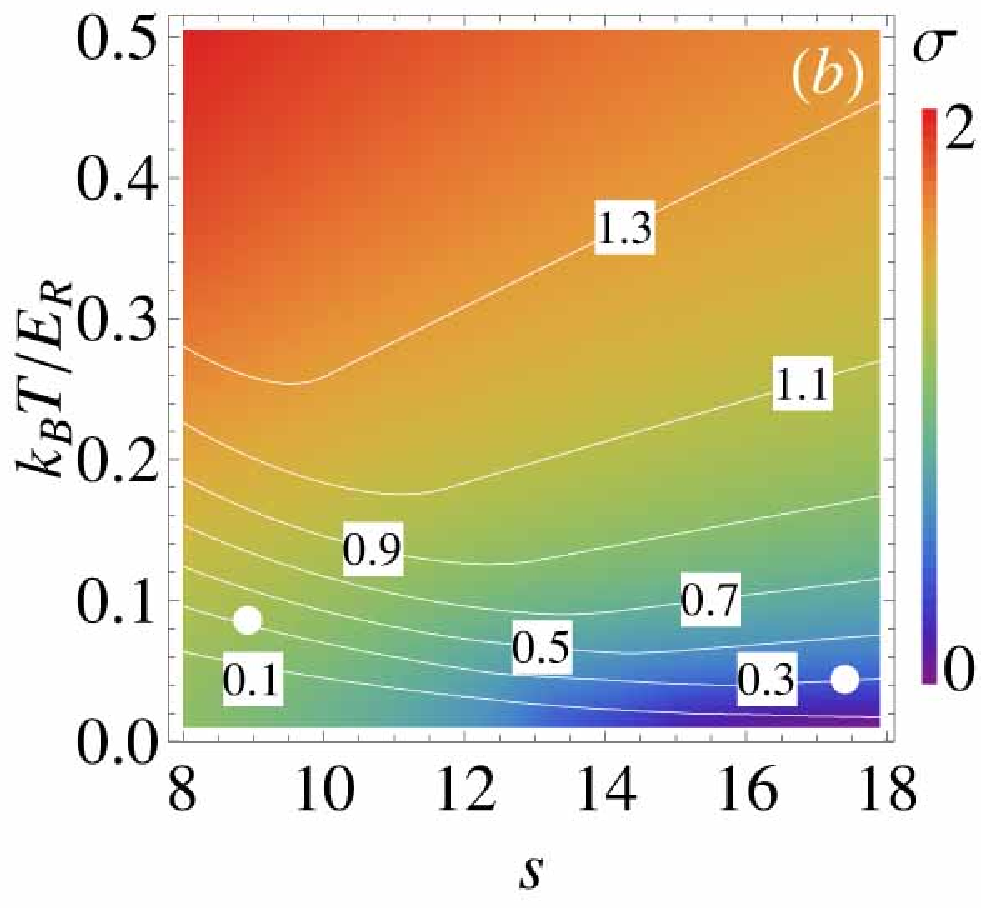}} \\
      \resizebox{45mm}{!}{\includegraphics{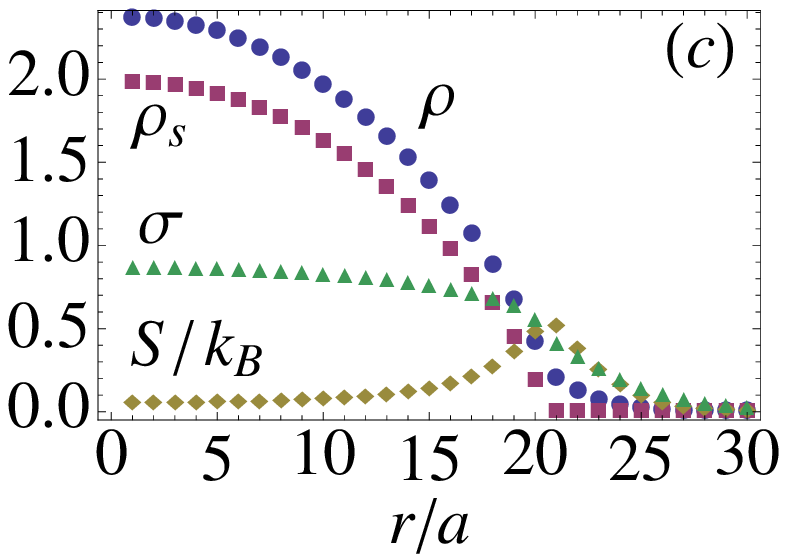}} &
      \resizebox{45mm}{!}{\includegraphics{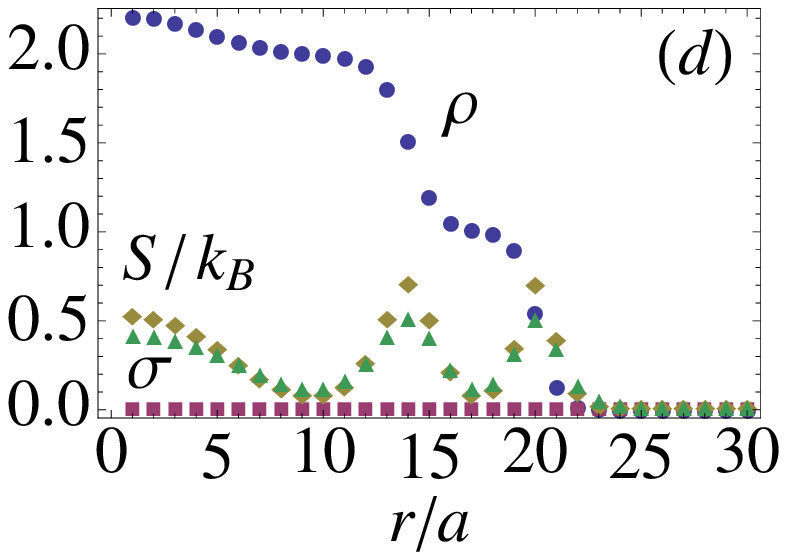}} \\\
    \end{tabular}
\caption{(Color online) (a) The isentropic curves and the density plot of
   the condensate
   density per particle. (b) The isoentropic
   curves and the density plot of the number fluctuation per
   particle. (c) and (d) Spatial distributions of $\rho$
   (circle, blue), $\rho _s$ (square, purple), $S/k_B$ (diamond, blown), and
   $\sigma$ (triangle, green) at $(s, k_BT/E_R)=(8.9,0.085),(17.4,0.045)$
   respectively. This two $(s,k_B T/E_R)$ are on the adiabatic line
   $S_{total}/N k_B = 0.3$, which are indicated by two white dots in (a) and
   (b). The condensate density exists in (c) while the wedding-cake
   structure is exbited in (d).}
    \label{AdiabaticForTrap}
  \end{center}
\end{figure}

Let us consider the system with the harmonic trapping potential $V_t (r) =
\frac{m}{2}\omega ^2 r^2$, which usually exists in experiments. Note
that $r$ is the distance from the center of the trapping potential. In the
presence of the trapping potential, the system becomes inhomogenous and
has the mixture of SF and N phases. In order to investigate the behavior of
the adiabatic heating/cooling in this case, we perform the same calculation as
the homogeneous case except inserting the local chemical potential $\mu (r) = \mu
_0 - V_t (r)$. We take the trapping frequency $\omega = 2 \pi
\times 24$, the lattice size to be $65^3$, and the total number
density $N = 2 \times 10^5$  as in the Greiner's work
\cite{Greiner02}. Figs.~{\ref{AdiabaticForTrap}} (a) and (b)
show qualitativley same results as homogeneous
case.
The adiabatic cooling occurs in the presence of condensate while
the adiabatic heating occurs when there is no condensate. Therefore we can
say that the behavior of adiabatic heating/cooling does not change
essentially 
whether there is the trapping potential or
not. We note, however, that the effects of the trapping potential is important
quantitatively for realizing the strong correlated system as shown in
Refs.~\cite{ho2007,Gerbier07}. In Figs.~{\ref{AdiabaticForTrap}} (c) and (d), we plot the spatial
distributions of the occupation number, the condensate density, the
entropy, and the number fluctuation along the adiabatic line
$S_{total}/N k_B = 0.3$ indicated by two white
dots in Figs.~{\ref{AdiabaticForTrap}} (a) and (b). We see the high condensate density in Fig.~{\ref{AdiabaticForTrap}} (c), while the wedding-cake
structure is exhibited in Fig.~{\ref{AdiabaticForTrap}} (d). Since we use local
density approximation, further detailed studies are needed for the case with
the trapping potential.

In conclusion, we calculated the three-dimensional Bose-Hubbard model at finite temperatures
within the mean-field approximation and investigated 
the effects of the adiabatic loading of optical lattices to the temperature. The
lattice-height dependence of the isentropic curves for given initial temperatures in case of
the homogeneous system was computed. We found that
the cooling occurs in SF phase while the heating occurs in N phase and
 the efficiency of the adibatic cooling/heating is higher for higher
 temperatures. Its behavior is determined by whether the dispersion
 relation of the system increases or decreases as loading optical lattices. Finally, we showed that the case with the trapping potential is
 essentially same as the homogeneous one.

$\textit{Note added.}$ Recently we have become aware of a related paper by
Pollet $\textit{et al.}$ \cite{pollet-2008}. They studied the
adiabatic loading in one-dimensional and two-dimensional optical
lattices by quantum monte carlo method. The effect of the trapping
potential to the adiabatic loading was investigated with high accuracy
in these lower dimensional cases.

\begin{acknowledgments}
S.Y. thanks to S.~Miyashita, N.~Kawashima, and I.~Danshita for fruitful
 discussions. S.Y. is supported by NAREGI Nanoscience
 Project from
 Ministry of Education Culture, Sports, Science, and Technology,
 Japan. S.K. is supproted by JSPS (Japan Society for the Promotion of
 Science) Research Fellowship for Young Scientists.
\end{acknowledgments}
\bibstyle{apsrev}

\end{document}